\documentclass{article}
\pdfoutput=1

\usepackage{arxiv}

\usepackage{algpseudocode}

\usepackage{array}

\usepackage{url}

\usepackage{comment}
\usepackage[inline]{enumitem}
\usepackage{lineno,hyperref}
\usepackage{float}
\usepackage{amsmath}
\usepackage{amssymb}
\usepackage[ruled,vlined]{algorithm2e}
\usepackage{multicol}

\usepackage[utf8]{inputenc} 
\usepackage[T1]{fontenc}    

\usepackage{amsfonts}       
\usepackage{nicefrac}       
\usepackage{microtype}      

\usepackage{graphicx}
\usepackage{natbib}
\usepackage{doi}

\title{``Is not the truth the truth?'’: Analyzing the Impact of User Validations for Bus In/Out Detection in Smartphone-based Surveys}

\author{Valentino~Servizi\\
	Department of Technology, \\
	Management and Economics\\
    Technical University of Denmark (DTU)\\
	\texttt{valse@dtu.dk} \\
	\AND
	Dan R.~Persson \\
	Department of Applied Mathematics \\
	and Computer Science\\
    DTU\\
	\And
	Francisco C.~Pereira \\
	Department of Technology, \\
	Management and Economics\\
    DTU\\
	\And
	Hannah~Villadsen \\
	Department of People and Technology\\ 
	Roskilde University\\
    Denmark\\
	\And
	Per~Bækgaard \\
	Department of Applied Mathematics \\
	and Computer Science\\
    DTU\\
    \And
	Inon~Peled\\
	Department of Technology, \\
	Management and Economics\\
    DTU\\
	\And
	Otto A.~Nielsen\\
	Department of Technology, \\
	Management and Economics\\
    DTU\\
}
\renewcommand{\shorttitle}{``Is not the truth the truth?''}

\hypersetup{
pdftitle={``Is not the truth the truth?''},
pdfsubject={Arxive},
pdfauthor={Valentino~Servizi et al.},
pdfkeywords={Ground-truth, D2D interactions, Autonomous vehicles, Bluetooth low energy, Internet of things},
}
\begin{document}
\maketitle

\begin{abstract}
Passenger flow allows the study of users’ behavior through the public network and assists in designing new facilities and services. This flow is observed through interactions between passengers and infrastructure. For this task, Bluetooth technology and smartphones represent the ideal solution. 
The latter component allows users’ identification, authentication, and billing, while the former allows short-range implicit interactions, device-to-device. To assess the potential of such a use case, we need to verify how robust Bluetooth signal and related machine learning (ML) classifiers are against the noise of realistic contexts. Therefore, we model binary passenger states with respect to a public vehicle, where one can either be-in or be- out (BIBO). The BIBO label identifies a fundamental building block of continuously-valued passenger flow. This paper describes the Human-Computer interaction experimental setting in a semi-controlled environment, which involves: two autonomous vehicles operating on two routes, serving three bus stops and eighteen users, as well as a proprietary smartphone-Bluetooth sensing platform. The resulting dataset includes multiple sensors’ measurements of the same event and two ground-truth levels, the first being validation by participants, the second by three video-cameras surveiling buses and track. We performed a Monte-Carlo simulation of labels-flip to emulate human errors in the labeling process, as is known to happen in smartphone surveys; next we used such flipped labels for supervised training of ML classifiers. The impact of errors on model performance bias can be large. Results show ML tolerance to label flips caused by human or machine errors up to 30\%.
\end{abstract}


\section{Introduction} \label{sec:introduction}
%
%
%
%
Passenger flow is a fundamental component for capacity estimation of public transport and for designing adequate infrastructure and services \citep{hankin1958a}. On bus transport, this flow measures passengers' variations in time and space, on the vehicles \citep{zhang2017a}. Multiple approaches promise real-time passenger flow estimation, but even the most advanced ones struggle with imprecision due to counting passengers indirectly, such as when paying by cash, or traveling without ticket \citep{zhang2017a}. Although autonomous buses and the internet of things (IoT) offer the opportunity of exploiting D2D (device to device) interactions for passenger flow beyond ticketing~\citep{wang2010a}, available solutions such as check-in/check-out (CICO), walk-in/walk-out(WIWO), or be-in/be-out (BIBO) \citep{Dekkers2007, Mezghani2008} all seem prone to errors. For example in the CICO case, using radio-frequency identification (RFID) technology for the interactions between smart-cards and readers, people often forget either the CI or CO action. In the WIWO case, multiple users can enter the same gate at the same time and confuse the counter. To contribute improving passengers' count accuracy and user experience in public transportation, we focus on enabling next generation BIBO for ticket-less trips. This application could allow passengers, for example, to pay with contact-less, radio-based identification, and communication via smartphone-Bluetooth without human intervention and without explicit interaction~\citep{narzt2015a}. This approach has the added advantage of being the most user-friendly for the growing population of smartphone-users, which is above $40\%$ worldwide and up to $80\%$ in western countries \citep{newzoo2020}, since it depends only on the user carrying his/her device as he/she would normally do. 

Although the global positioning system (GPS) is one of the most reliable and adopted technologies for outdoor tracking~\citep{servizi2020a}, GPS shows important limitations in urban areas~\citep{cui2003a}. The specific radio-signal frequency requires line of sight between sender and receiver, thus being affected by reflections from tall buildings and clouds. Similarly, Bluetooth is one of the principal technologies for proximity detection~\citep{Sapiezynski2017} applied to indoor tracking, and the specific radio-signal frequency brings other limitations. For example, a smartphone-based travel survey on the Silver Line bus rapid transit in Boston, Massachusetts, deployed BIBO technology~\citep{li2017a}, as a context detection system for a service quality survey, and so avoided collecting ground-truth, in form of labels, from surveyed passengers; D2D implicit interaction between smartphones and Bluetooth devices installed on buses verified passengers' presence aboard, independently from GPS sensors. In the same study, the authors expose cases where successful BIBO verification via GPS otherwise failed via Bluetooth, because smartphones could not receive Bluetooth signal within the bus, probably due to human body impedance relative to smartphone and Bluetooth device position. While a large body of literature presents a successful case for Bluetooth as indoor positioning technology, no previous work that we are aware of analyses in detail its use as independent measurement for labels and the impact labeling errors on the BIBO classifier performance. In this use case, Bluetooth reception errors might present themselves as flipping- and outlying-labels~\citep{rolnick2018a}, negatively impacting ML training and magnifying misclassifications.

Flipping-labels are known as items that human or machine classifiers labeled with a wrong class, despite the true one existing in the dataset; outlying-labels are items that belong to none of the classes in the dataset, but were mistakenly labeled as one of these classes~\citep{sukhbaatar2015training}. The impact of these two problems on ML classifiers is extensively studied for independent and identically distributed (IID) datasets~\citep{ahmed2020a, sukhbaatar2015training, hendrycks2018a, natarajan2016a, yi2019a, rolnick2018a, nettleton2010a, rodrigues2017a, liu2019a, jiang2020a},  such as for images, but not for time-series, such as Bluetooth or space-time GPS trajectories.

To bridge the gap between the limitations mentioned in the preceding two paragraphs, we conducted a case study on a BIBO, smartphone and Bluetooth Low Energy (BLE) based system, with the following research questions:
\begin{enumerate}
    \item During the ground-truth collection, what is the users' response to wrong labels?
    \item After ground-truth collection, what is the ML classification performance based on various features extracted from BLE, GPS and accelerometer?
    \item What is the resilience of ML supervised methods to flipping-labels?
\end{enumerate}

\begin{figure}[!ht]
\centering
\includegraphics[width=\textwidth]{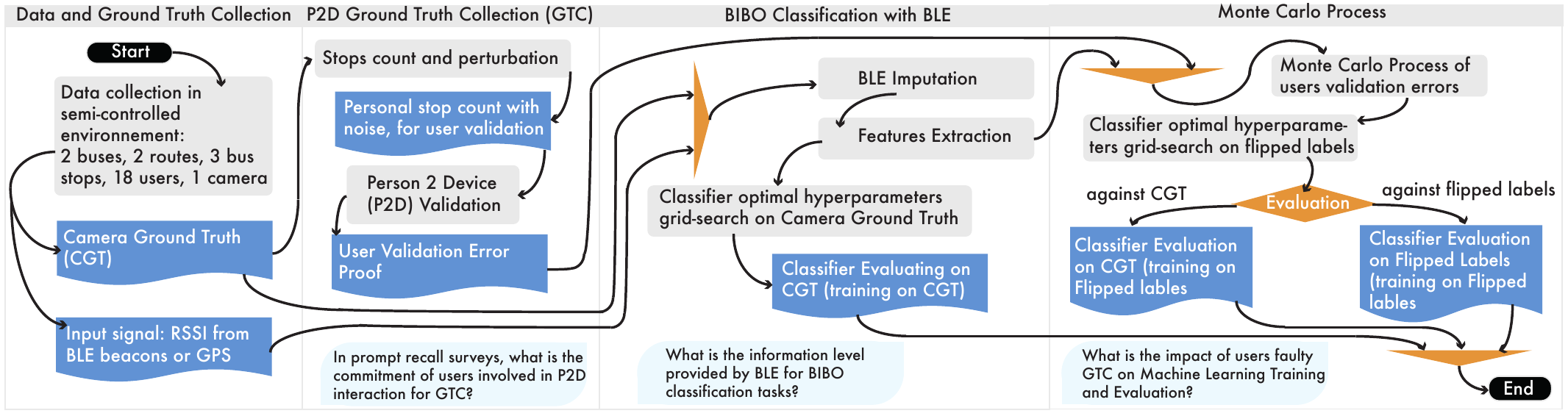}
\caption{Experiment workflow.}
\label{fig:first_case}
\end{figure}

Fig. \ref{fig:first_case} shows the process we executed through the following steps. First, we designed and implemented a smartphone-BLE platform. Second, we set up and ran an experiment involving a simple transport network composed of two autonomous buses operating on two routes and three bus stops, with a BLE device on each bus and bus stop. Third, we involved eighteen users and we video-recorded each of their trips through this network; simultaneously, users' smartphones native (Android and iOS) application programming interface (API) read BLE devices' signal strength and classified the transport mode from the time-series of the inertial navigation system (INS). Our proprietary application stored these trajectories on a database. Fourth, we labeled the trajectories using video recordings as ground-truth with BIBO binary labels. Fifth, we created a Monte Carlo (MC) process to simulate labeling errors, i.e. flipping-labels, with various noise levels on recorded trip time-series. Finally, to understand the error tolerance, we evaluated and compared multiple classifiers, both on true and noisy labels.


The experimental setting incorporates multiple real-world conditions typical of urban high density contexts: overlapping BLE fields, multiple makes and types of smartphones, native applications for the two main operating systems (OS), bus switching routes, bus moving at low speed, subjective preferences on how and where users carry their smartphones, or where they stand, both while traveling on bus and waiting at the bus stop. In such a BIBO system setup, we yield results suggesting that BLE signal alone is robust to labeling errors, and performs significantly better than commercially-available classifiers based on INS.

\section{Related Work} \label{sec:relatedworkI}
This section focuses on two main bodies of literature that contribute to expose perspectives relevant for this use case: one on deployment of BLE beacons networks and signal processing for location prediction and activity classification, the other on the problem of label noise for ML classifiers. This section pinpoints candidate parameters and methods considered for designing the experimental setup of this work.

\subsection{Bluetooth applications}
\label{sec:relatedWork}
BLE stems from Bluetooth and WiFi protocols, and specializes in IoT applications; the communication is one-to-many, involves few bits of data to be broadcast frequently, and requires no pairing operation with other devices. All these properties make BLE technology particularly suitable for proximity detection~\citep{servizi2020a}. Although in some field is heavily unbalanced towards other sensors, such as GPS and INS~\citep{servizi2020a}, BLE and WiFi are considered promising technologies even for transport mode detection in complex multimodal transport chains~\citep{brouwers2013a, muthukrishnan2009a, y2011a, gonzalez2010a, wind2016a, bjerreNielsen2020a}. For example, \cite{bjerreNielsen2020a} perform transport mode detection based on received signal strength (RSSI) from Wi-Fi and Bluetooth signals, measured in decibels. The study analyzes and compares three supervised classifiers: random forest, logistic regression, and support vector machines. None of these methods involves artificial neural networks.

For implicit BIBO classification, \cite{narzt2015a} propose an architecture where Bluetooth receivers are inside the bus while passengers carry a BLE device. The study carries out several experiments to recreate bus-space realistic conditions and analyses multiple configurations for Bluetooth receivers and device positions. No real users nor vehicles are involved in the study. However, the conclusion cautiously supports the hypothesis that larger-scale deployment of such a system is feasible. To further investigate potential interactions with the environment, the study highlights the need for a survey under realistic conditions from a larger-scale deployment perspective.

An independent, complementary, and substantial body of literature focuses on multiple sensors and algorithms for Mobile Anchor Node Assisted Localization \citep{han2016a}, where WiFi and BLE signals are extensively studied in general, and in particular for indoor tracking \citep{yassin2017a}. Among the methods available, geometric approaches are widespread, e.g., based on the Friis equation \citep{kotanen2003a}, and trilateration \citep{subhan2011a}. These methods rest on the knowledge of each device position and radio-signal propagation physics to approximate a receiver's location based on reception strength.
Prevalent RSSI fingerprints approaches are ML-based, e.g., on k-nearest-neighbor and Kalman-Filters (KL) ~\citep{chen2015a, subhan2013a, p2012a}. These algorithms rely on mapping a geo-spatial context with a sample of signal-strength-records, received from the devices on the range; grid resolution on the mapped space and signal-sample-size depend on the location accuracy required by the use case.

A natural extension of these technologies in the field of intelligent transport systems, is the study of vehicles to anything (V2X) communication. Whereas Bluetooth in general is not considered optimal for bi-directional communication due to slow paring process \citep{moubayed2020a}, BLE technology is substantially different and is able to trigger events in smartphones' OS, without any paring operation \citep{AppleBLE2021}.

In summation, the above works indicate several pre-requisites for successful BLE application:
\begin{enumerate*}[label=(\roman*)]
    \item BLE signal transmission rate above 0.3 Hz;
    \item The density of the Bluetooth beacons network above one device every 30 square meters; 
    \item Appropriate imputation of RSSI readings.
\end{enumerate*}

\subsection{Noisy labels in machine learning classifiers}
The problem of noisy data receives a lot of attention from the research community. The cause of noise in labels is manifold and use case dependent. For example, crowd-sourced labeling of images relies on expertise and attention of labelers, which they may not always have~\citep{rodrigues2017a}. Similarly, in prompted recall surveys, users validate travel diaries with different dedication levels, and may therefore, negatively affect the quality of what is often perceived as ground-truth~\citep{servizi2020a}. Consequently, noisy labels in turn negatively affect the classification accuracy of supervised or semi-supervised ML methods, which depend on these labels in the training process. 

Previous systematic studies on noisy labels compare multiple supervised classifiers on multiple synthetic datasets~\citep{nettleton2010a}, and analyze how robust learning algorithms are to noise~\citep{Beigman2009, Manwani2013, man2015a}. Another research line works on noise cleansing or labels correction methods~\citep{Barandela2000, brodley1999a, miranda2009a}. Numerous alternative approaches exist for improving classification accuracy in the presence of noisy labels, for example: 
\begin{enumerate*}[label=(\roman*)]
    \item To pinpoint wrong labels, majority voting across multiple neural networks~\citep{Yuan2018}.
    \item To learn labels' noise distribution, specialized layers for artificial neural networks~\citep{rodrigues2017a, sukhbaatar2015a, Jindal2016}.
    \item To predict the noise affecting training, conditional noise models~\citep{xiao2015a}.
    \item To reduce the number of labels necessary for training, semi-supervised approach achieved with generative models~\citep{kingma2014a}.
    \item To leverage on existing high-quality sub-set of labels, propagation methods of these labels \citep{zhu2009a}.
    \item To learn labels on the fly and reduce human errors, graph-based label propagation methods~\citep{Fergus2010}.
\end{enumerate*}

However, despite the wide body of previous work, our use case did not receive enough attention, and thus no conclusions can be drawn as of its potential.
Whereas in existing work datasets are exclusively IID, we investigate on dependent observations over time. Further, we explore the impact of varying labels' quality on model performance, presenting a real-world dataset with high-quality ground-truth, and leveraging Monte Carlo simulations for labels' variation study. In contrast, existing works rely on synthetic datasets, which on one hand offer reliable ground-truth, but on the other hand yield biased measurements compared to real data.
In addition, we try to answer the following two questions:
\begin{enumerate*}[label=(\roman*)]
\item What is the commitment of users involved in a person to device (P2D) explicit interaction for ground-truth collection? 
\item What is the information level that BLE provides for BIBO classification tasks with optimal ground-truth?
\end{enumerate*}



\section{Methods and materials}\label{sec:methods}
\begin{figure}[!ht]
\centering
\includegraphics[width=0.7\textwidth]{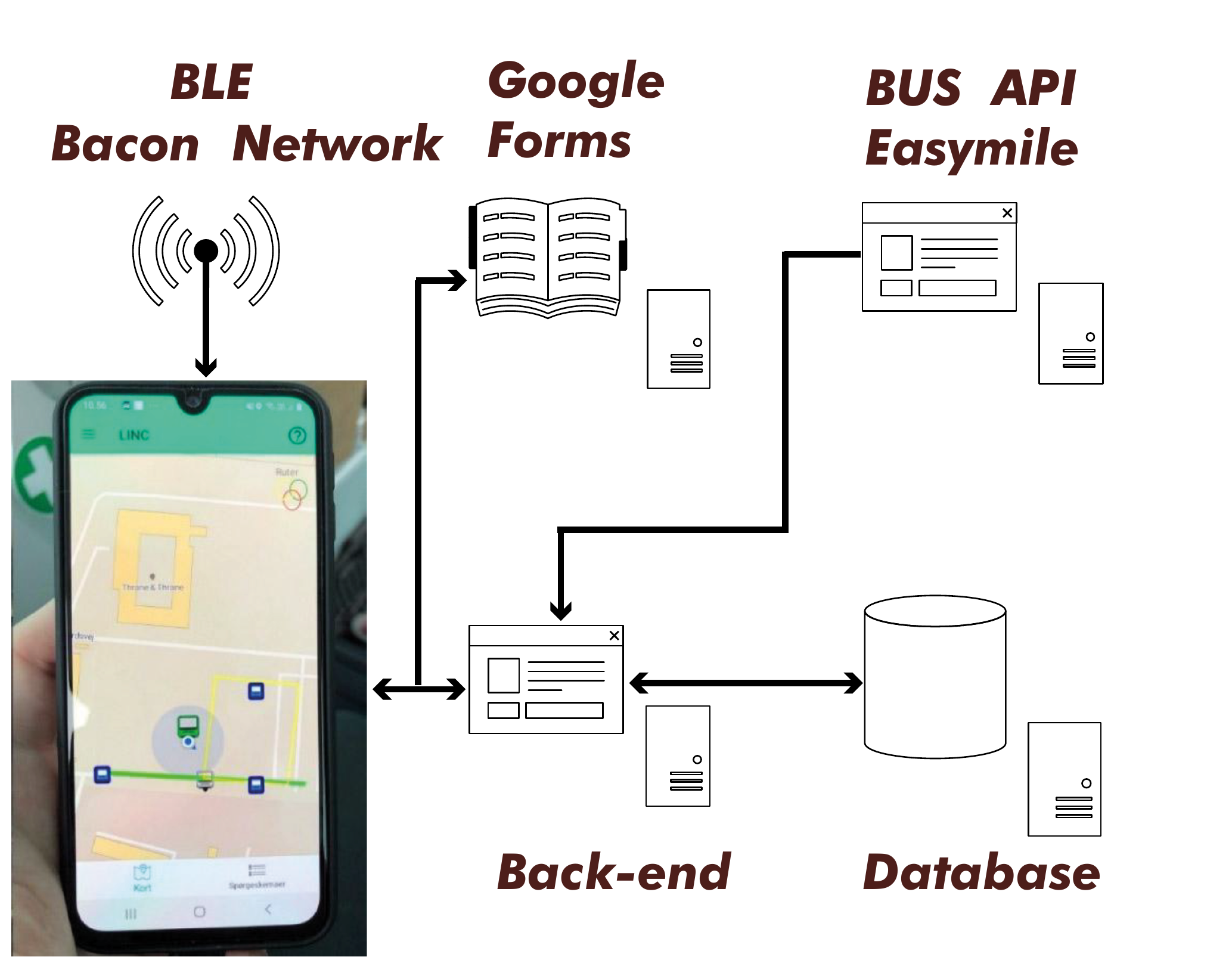}%
\caption{Sensing platform.}
\label{fig:sensingArchitecture}
\end{figure}

To assess BIBO error tolerance under the experiment setup, first we need to understand how users collect faulty ground-truth, and then use this knowledge to derive a Monte Carlo process generating the same noise on the labels. Next, we can provide a broader analysis over the impact of noisy labels on Machine Learning training and evaluation steps, and mostly we can carry out this analysis on real trajectories. Fig. \ref{fig:first_case} describes the methodological process we adopted; Figure \ref{fig:sensingArchitecture} presents the BIBO platform we designed, implemented and deployed for data collection.

\subsection{Sensing platform}
The smartphone sensing platform's main components are the front-end applications and back-end. The front-end is specific, or native, for Android and iOS. The apps contain the following features: data collection from onboard sensors and native APIs, such as users’ transport activities classification; data transfer to the back-end from a local buffer that avoids data loss in case of external connectivity problems; and lastly, real-time tracking of buses on a map, with bus stops. 
The sensors we target are GPS, 
and BLE signal strength perceived from the BLE beacons network. 
To improve smartphones' battery efficiency, we monitor users' activities. We switch on and off smartphone GPS and data transfer, when the user is active and inactive, while the accelerometer detects the state. 
The back-end, which includes several specialized APIs, is responsible for exposing the information from the autonomous vehicles to the smartphone and handing the data from smartphones to the database.

As opposed to another architecture presented for BIBO ~\citep{narzt2015a}, where users carry BLE devices, and buses are equipped with signal processing devices, we decide to follow the architecture of smartphone-based travel surveys~\citep{servizi2020a}, where users carry their smartphone, which is also a signal processing device, while buses are equipped with BLE devices. This configuration presents two main advantages. First, we extend the beacons' network outside the bus, at the bus stops. Second, we allow users to carry their phones and not interfere with their normal behavior while picking up signal from multiple sources. From the perspective of a larger-scale deployment, the experimental setup seems more realistic under these two conditions. The installation cost of a beacon device should be a fraction of the Raspberry-Pi deployed in ~\citep{narzt2015a}; furthermore, users should carry their smartphones only and not a new device. This last element introduces a new random variable: The varying quality of sensors installed in different smartphone models, and the sample collected in this experiment is not representative of this broad population. Thus, to collect consistent data, we rely on the standardization of sensors and protocols represented on the aforementioned OS. 
We also collect the accelerometer-based activity recognition that these OS offer via APIs to discriminate between states, such as: automotive, bicycling, walking, running, stationary and unknown~\citep{AppleAPI2021, AndroidAPI2021}.
 
\subsection{Experiment setup and data}\label{sec:experimentSetup}

The possibility of replicating beacons' density of indoor settings, which should be $>\frac{1}{30 \ m^2}$, is not realistic from this use case's scale-up perspective (see Sec. \ref{sec:relatedWork}). However, installing devices both on the bus and on bus stops, which is more  realistic, allows a temporary and nearly-optimal density of the beacons' network, at least between passengers' boarding and alighting, and when the bus stations in front of a bus stop.

The setup consisted of two autonomous vehicles operational on two distinct routes and three bus stops. To allow passengers' transfer between the buses, one bus-stop was shared between the routes, additionally sharing a segment of the test track; during the experiment, the buses' assignment to the route has been switched for technical problems, similar to real world settings. The BLE beacon network counted one device per bus and one device per bus stop, with five devices in total (see Fig. \ref{fig:sensingArchitecture}). Each device transmitted at the rate of $1.667 \ Hz$ and $-8 \ dBm$ power. As these two settings affect the battery life of BLE beacons, the decision considers a realistic battery life expectation above one year, within the frequency recommendations from indoor studies (see Sec. ~\ref{sec:relatedWork}).

Smartphone onboard sensors collected trajectories for twelve users only, for a total of $13,723  \ points$. We stored for each of these timestamps: GPS longitude and latitude; 
5 RSSI readings from the BLE beacons network, one for each device, and transport-mode as classified by off-the-shelf accelerometer-based classifiers available on both Android and iOS operating system. The high number of unavailable trajectories, approximately $\frac{1}{3}$ of the total, is the result of two distinct problems. Four users did not grant the permission to access location sensors, resulting in no database records.

To count passengers' flow, we installed a high-resolution video-camera pointing to the buses' doors at each bus stop as the principal ground-truth. However, the three cameras in combination also allowed the full surveillance of the track. A problem with the video-cameras, prevented the determination of high quality ground-truth for three users. Using video footage as ground-truth for the trajectories successfully collected from the remaining users, we provided a set of binary labels consistent with the BIBO model~\citep{narzt2015a}, on each point: inside or outside the bus.

To collect feedback from users after the experiment and link each user's feedback to the other data collected from smartphones, we rely on electronic forms with a pre-filled unique identifier corresponding to the user.


\subsubsection{Procedure}
We distributed a paper-based form to each user. The form included the experiment description and the information on data collection and use exclusive for research purposes (GDPR complainant). Then we briefed each participant on the following steps: 
\begin{enumerate*}[label=(\roman*)]
\item Install on smartphones the application we published to the application stores (for beta testing). 
\item Read general conditions and grant the application permission to access smartphone sensors and activity recognition. The latter performs transport mode detection \citep{AndroidAPI2021, AppleAPI2021}. 
\item Wear a sleeve number to ease the ground-truth collection from video recordings.
\item Use the transport network with the commitment to enter and exit the bus more than once, and with the possibility of walking between bus stops.

\item Count the total number stops, defined as the discontinuities between transportation modes, and expect a message stating our count, in the following days, with the request to validate or correct such a count.
\end{enumerate*}

Finally, we answered any questions raised by the participants, and from all the participants willing to participate we collected a paper-based signed authorization to proceed with experiment and data collection \footnote{This project is a social science study, includes data and numbers only, is not a health science project, and does not include human biological material nor medical devices. Consequently, in Denmark, where the data collection took place, the Health Research Ethics Act provides a dispensation for notification to any research ethics committee.}.

\subsubsection{Participants}
Active ground-truth collection P2D, which users provided in the days following the experiment, included fourteen valid replies. We counted the total number of stops for each user from video-recordings. To explore the users' commitment and the quality of a P2D ground-truth collection, we introduced a level of noise in these counts before submitting the validation request, on a random sample of users. For the noise distribution, we assumed that validation errors could be Poisson distributed, similarly to OD matrix counts~\citep{cascetta1988a}, as each error event is discrete and has minimal probability. 

\subsection{Wizard of Oz (WoZ) for P2D ground-truth collection} \label{sec:wizardOfOz}
In this case, WoZ refers to the experimenter pretending that a BIBO system is operational on the test-bed \citep{riek2012a}. The role of the user is to validate the measurements of such a BIBO system. Therefore, the user is briefed to count how many times he or she alighted from or mounted on a bus. To observe the P2D validation dynamic, the experimenter then provides WoZ's count to the user. In particular, as stated in Sec. \ref{sec:experimentSetup}, we assume that a Poisson distributed random error affects users' count. Since users' validation seems to support this hypothesis, we use this distribution to simulate users' validation errors within the Monte Carlo simulation described in Sec. \ref{sec:montecarlo}, where counts errors propagate to the time-series' labels, flipping a BI in BO, or vice versa. 

\subsection{Data preparation and classifiers for BIBO}
To assess BLE beacons' signal performance in determining users' presence inside or outside the buses, we use GPS as one of the benchmark. 
From GPS we extract the following features: distance between points, bearing, and speed~\citep{servizi2020b, Dabiri2018}, which we process as time series extracting the same features extracted for BLE beacons. Table \ref{tab:fearCorr} presents the list of features collected in 10 seconds moving window. 
Further, from smartphones' OS we collect the binary classification automotive vs. everything else, compatible with BIBO in this context, which is based at least on accelerometer.
Thus, we rely on the following tools, which we apply separately to BLE and GPS signal.

\subsubsection{Framework}
Scikit-learn is a popular python-based framework that includes several effective ML models. Random Forests (RF) represent a reliable and scalable supervised method for this task~\citep{Zhou2019}. At the same time, Multi-layer perceptron (MLP) can be considered a building block of generative models, which can operate semi-supervised or unsupervised~\citep{kingma2014a}. Therefore, we include these two supervised classifiers in the study. The following sections present further details, on preparation, training, and validation of the classifiers.

\subsubsection{Random Forest}
RF evolve from decision tree predictors, averaging results from multiple of these predictors. The effect is a more accurate classifier less prone to over-fitting. The training phase starts with bootstrapping~\citep{Breiman2002}, which consists of several sub-samples with replacement from the training dataset. Each training sub-set is then split into in-bag~\citep{Breiman2002} (IB) and out-of-bag~\citep{Breiman2002} (OOB). The latter's size is one-third of such a sub-set, while the former accounts for the rest. A decision tree is constructed from each IB, while the attributes are sampled randomly to determine the decision split~\citep{Breiman2002}. Finally, the RF output is aggregated over all individual trees, and the output is the class with the highest average probability, whereas in classical majority voting the output is the most common class prediction among trees.

\subsubsection{Multi-layer perceptron}
Perhaps we can consider it the most simple feed-forward artificial neural network~\citep{Bishop2006}. MLP incorporates multiple layers for logistic regression. Multiple perceptrons, or neurons, compose each layer and handle 
nonlinearities through activation functions, such as sigmoids and rectified linear units (ReLU). For classification, each neuron's weight and bias is trained by minimizing the cross-entropy between the class predicted by the network and the ground-truth. These parameters are iteratively updated at each classification attempt, defined epoch, by back-propagating the resulting stochastic gradient towards the cross-entropy local minimum.

Training artificial neural networks requires large datasets. In this case, it is arguable whether the dataset size is appropriate or not. However, given the possible future extension of this experimental setup to real life operations, we are interested in investigating the potential.

\subsubsection{Hyperparameters grid search}
To perform this task we used \textit{GridSearchCV}, a specialized library available in Sklearn. To obtain a set of optimal hyperparameters, we perform a 5-fold cross-validation on the training-set exploring those that Table \ref{tab:RFparametersSet} and Table \ref{tab:MLPparametersSet} describe for RF and MLP. In a following step we train the classifier on the training-set, fixing these optimal hyperparameters, and we perform the evaluation on the test-set. Sec. \ref{sec:montecarlo} provides further details on this process within the  simulation of ground-truth collection errors causing flipping-labels.
\vspace{-5pt}
\begin{table}[ht]
    \centering
    \caption{Random forest hyperparameters search space.}
    \begin{tabular*}{\textwidth}{r|l}
         \hline 
         Number of estimators & $\in \{10, 20, 100, 200, 500\}$ \\
         Max features & $\in \{\text{auto}, \text{sqrt}, \text{log2}\}$\\
         Max depth & $\in \{3, 4, 6, 7, 8\}$\\
         Criterion & $\in \{\text{gini}, \text{entropy}\}$ \\
         \hline
    \end{tabular*}

    \label{tab:RFparametersSet}
\end{table}
\vspace{-5pt}
\begin{table}[!ht]
    \centering
    \caption{Multi layer perceptron hyperparameters search space.}
    \begin{tabular*}{\textwidth}{r|l}
         \hline 
         Hidden layers/sizes & $\in \{1 \ \text{layer} \ (L)\rightarrow[50 \ \text{neurons} \ (N)], $\\
         &$3L\rightarrow[10N, 50N,10N], $\\
         &$4L\rightarrow[10N,50N,50N,10N]\}$ \\
         Learning rate strategy& $\in \{\text{constant}, \text{invscaling}\}$\\
         Learning rate coefficient & $\in \{10^{-2}, 10^{-3}\}$\\
         Activation funcions & $\in \{\text{\text{ReLU}}\}$ \\
         Optimizer & $\in \{\text{adam}\}$ \\
         \hline
    \end{tabular*}

    \label{tab:MLPparametersSet}
\end{table}

\subsubsection{Validation process}
The risk of information spill-over between training- and validation-set is higher when working with time-series. \citep{Hillel2020} shows that the violation of the out-of-sample (OOS) principle is not rare in the existing literature. Such a violation yields a virtual higher performance when evaluating a classifier, resulting in a biased measurement. Even in the assumption of non-violation of the OOS principle, researchers have several options for assessing a classifier, such as hold-out, leave-one-out, and cross-validation. 
\begin{enumerate*}[label=(\roman*)]
\item \label{item:1} In the hold-out case, typically, the training-set should use approximately $\frac{2}{3}$ of the dataset; the validation-set, the remaining $\frac{1}{3}$. Training and validation proceed only once and yield the model performance based on the sole validation-set.
\item \label{item:2} In the leave-one-out case, the training-set should use a dataset's random sample of size $M-1$, where M is the dataset's cardinality; the validation-set, the remaining one sample. Training and validation proceed M times and yield the model performance as a distribution over M-validations.
\item \label{item:3} In the cross-validation case, the dataset is split into N equal partitions; the training-set uses N-1 partitions, while the validation-set uses the remaining one partition. Training and validation proceed N times and yield the model performance as a distribution over N-validations.
\end{enumerate*}
The approach \ref{item:1} is computationally light-weight, but the resulting performance estimation might be negatively biased; \ref{item:2} is unbiased but could present a large performance variance, and the method is computationally expensive; \ref{item:3} is a good compromise between the previous two~\citep{baraldi2005a}. Sec. ~\ref{sec:montecarlo} explains how our simulation combines these three methods with the hyperparameters grid search to provide an optimal and unbiased performance estimation and how we sample training- and validation-set to avoid OOS violation.

\subsubsection{Validation metrics}
As performance estimation metrics for binary classifiers, the literature presents a broad use of precision \eqref{eq:precision}, recall \eqref{eq:recall}, F1-score \eqref{eq:f1-score} and accuracy \eqref{eq:accuracy}.  Although these metrics are often sufficient, we introduce the measure of the area under the receiver operating characteristic curve (AUC). This curve describes the true-positive-rate (TPR) \eqref{eq:recall}, which is another identification for the recall, as a function of the false-positive-rate (FPR) \eqref{eq:ROC-FPR}, within the domain of any possible FPR  $\in [0,1]$.
We can derive these metrics directly from the confusion matrix, i.e., true positives ($T_p$), true negatives ($T_n$), false positives ($F_p$), and false negatives ($F_n$). 

The binary BIBO classes are quite imbalanced and the classification task is rather challenging given the experiment's realistic conditions. We recreate a congested urban context with multiple buses operating at speed similar to walking pace, in proximity to various bus stops. Whereas F1-score identifies cases where the random classifier is better than our classifier, AUC identifies also cases where the classifier only predicts the larger class. The domain of precision \eqref{eq:precision}, recall \eqref{eq:recall}, F1-score \eqref{eq:f1-score}, and accuracy \eqref{eq:accuracy} is $\in [0, 1]$, the higher the value the better. 
AUC's domain is also $\in[0,1]$. The interpretation of AUC coefficient for random classifiers results in the same distribution of the F1-score, strictly around $0.5$. In contrast with F1-score, for cases where classifiers predict only one class AUC presents the same distribution of the random classifier. Therefore, with AUC we expect good classifiers above $0.5$ threshold, with higher values being better. Below this threshold a classifier would be consistent in predicting the wrong class. Both random and trivial classifiers should score $0.5$ AUC in average.
To assess our simulation results against both the random classifier and the single-class-predictor, AUC measures how well predictions are ranked, and is invariant to scale and classification-threshold~\citep{cantarero1996a}. 
Since at this stage we are agnostic on the cost of false positives and false negatives, these two properties are not a disadvantage, as opposed to the advantages in assessing the classification performance with different levels of errors on the labels, over a large number of samples.
\begin{equation}
\label{eq:precision}
    P=\frac{T_p}{T_p+F_p}
\end{equation}

\begin{equation}
\label{eq:recall}
    TPR = R = \frac{T_p}{T_p+F_n}
\end{equation}

\begin{equation}
\label{eq:f1-score}
    \text{F1}=2\frac{P \cdot R}{P+R}
\end{equation}

\begin{equation}
\label{eq:accuracy}
    \text{A}=\frac{T_p + T_n}{Total \ population}
\end{equation}

\begin{equation}
\label{eq:ROC-FPR}
    FPR = \frac{F_p}{T_n+F_p} 
\end{equation}

\subsection{Simulation of Error Distributions}\label{sec:montecarlo}
After data preparation, as summarized in Alg. \ref{alg:prepData}, we can proceed with the simulation (MS), as detailed in Alg. \ref{alg:montecarlo}.  In contrast to the literature studying flipping labels' problem using true ground-truth from a synthetic generation of datasets, we apply the following principles:
We use high-quality ground-truth from a realistic setup involving real vehicles, devices, and people, producing real time-series from BLE and GPS sensors.
We employ the  method to simulate human errors as verified in the WoZ part of this experiment (see Sec. \ref{sec:experimentSetup} and \ref{sec:wizardOfOz} for methodology and Sec. \ref{sec:results} for experimental evidence).
We propagate such errors to the labels, assuming a state-of-the-art P2D ground-truth collection process, the same as real-world smartphone-based travel surveys.

Through the repeated sampling of user-by-user, the number of errors per user, and consequent propagation on the labels of each trip, at each repetition we train and evaluate ML classifiers against the random classifier, over features extracted from BLE sensors, versus features extracted from GPS sensors. 
We ensure the OOS validation principle on both grid-search search and methods' evaluation by randomly sampling 20\% of the users and then picking all their trajectories to compose the validation set. Thus, we take the complement for the training-set.

We yield performance's unbiased estimation by applying a hold-out scheme within each run, where the training partition allows a grid-search through 5-fold cross-validation. Since the validation-set evaluation runs multiple times, one for each draw, the results we obtain are comparable to a leave-one-out scheme or better, rather than the hold-out scheme. 
This process does not just flip labels, but also simulates the validation type of error in the experimental context (see Sec. \ref{sec:wizardOfOz}). For example, if the user does not validate location and count of his or her alighting, we flip the BO labels of the corresponding trip-leg, to match the BI label of the previous trip-leg. However, in the simulations, we also apply random flip of labels from BI to BO and vice-versa, as possible sanity check algorithms on the labels are not in the scope if this work. 

\section{Results} \label{sec:results}
In this sections we organize the results according to the research questions listed in Sec. \ref{sec:introduction}.
\subsection{People errors during ground-truth collection}
The experiment included video recordings of the ground-truth for eighteen users in total, that we used for the P2D validation experiment. 
The resulting confusion matrix on error distributions for labels (see Table \ref{tab:P2Dresults}), shows that $50\%$ of the received replies were perturbed. Nearly $60\%$ of the user modified the counts, while the remaining population confirmed the counts as received. One user confirmed the perturbed count, and two users modified the correct counts. Overall, more than $40\%$ of the validations contained at least one error, with average 0.7. 

\begin{table}[!ht]
    \centering
    \caption{Person to device ground-truth validation.}
    \vspace{5mm}
    \begin{tabular}{l|cc|cc|l}

 & Modified     & Confirmed     & Correct     & Wrong    &    \\
\hline
\multicolumn{1}{r|}{Perturbed}     & 6          & 1         & 3           & 4        & 7 \\
\multicolumn{1}{r|}{Not Perturbed} & 2          & 5         & 5           & 2        & 7 \\
\hline
& 8          & 6         & 8           & 6        &  14  \\
\hline
& \multicolumn{2}{c|}{14} & \multicolumn{2}{c|}{14} &   

    \end{tabular}

    \label{tab:P2Dresults}
\end{table}

\subsection{BIBO Classification Performance}
Results show that the difference between the random classifier and the accelerometer-based activity recognition is minimal. Random forest trained and evaluated with camera-ground-truth performs significantly better when classifying BLE beacons or GPS. In contrast, when processing features extracted from BLE, multi-layer perceptron performs worse than the random classifier in the same conditions.
Overall, the low performance of production-level classifiers based on accelerometer reflects this challenging and realistic experiment setup. Although we only simulated an online classifier, and our output was off-line in practice, the BLE signal shows potential for the BIBO task. GPS yields the highest accuracy, as expected, but at the most expensive battery cost, compared to both accelerometer and BLE \citep{servizi2020a}.

\subsection{BIBO Resilience to Label Flipping}
\begin{figure}
\centering
\begin{minipage}{.8\textwidth}
  \centering
  \includegraphics[width=\textwidth]{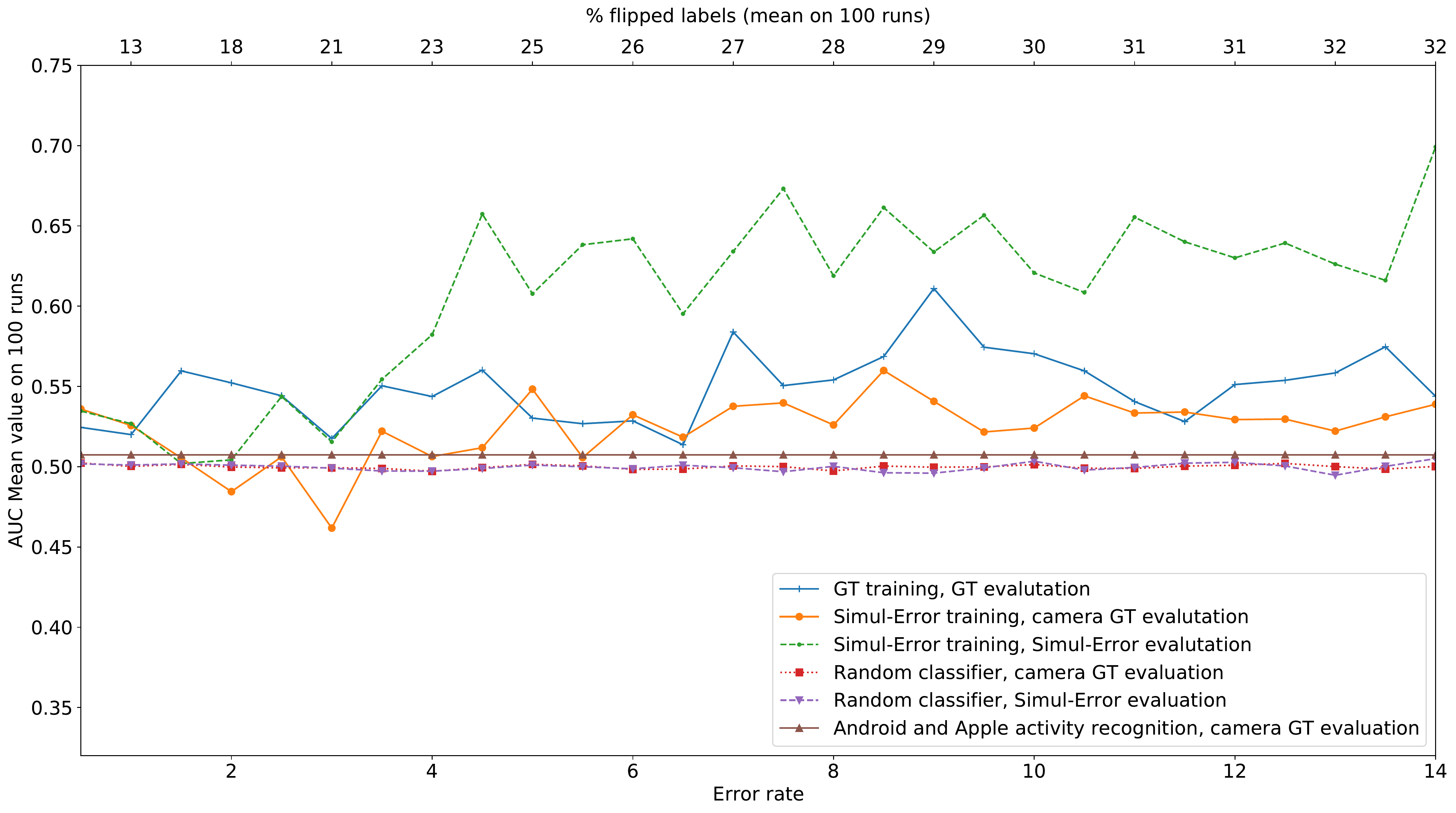}
  \vspace{-3mm}
  \caption{Random Forest one flip experiment 
  \newline AUC Classification task using BLE RSSI signal only (p-values $<< 0.01$)
  }
  
  \label{fig:AUCbleRFixxOF}
\end{minipage}%

\begin{minipage}{.8\textwidth}
  \centering
  \includegraphics[width=\textwidth]{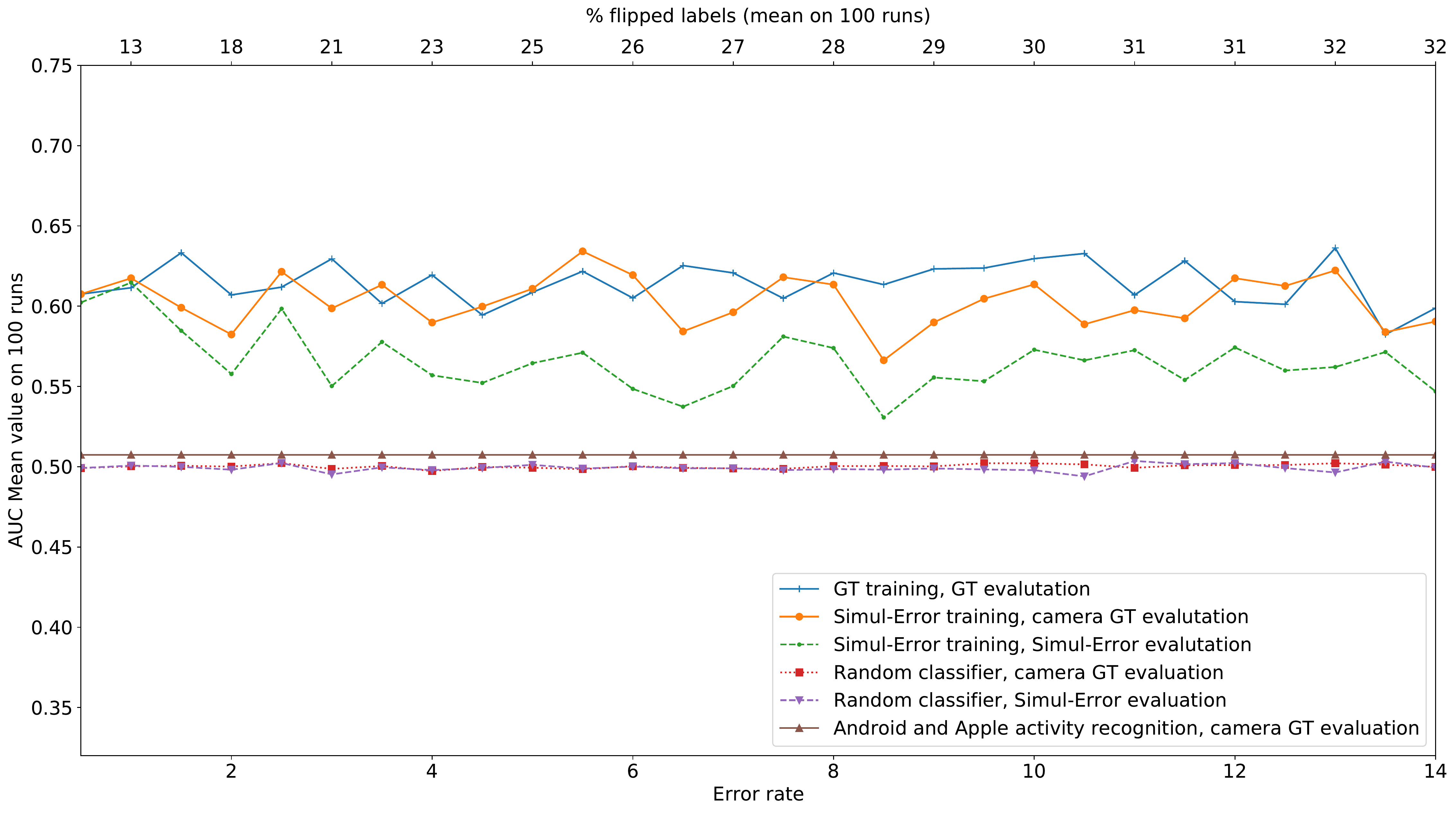}%
  \vspace{-3mm}
  \caption{Random Forest one flip experiment 
  \newline
  AUC Classification task using GPS signal only (p-values $<< 0.01$)
  }
  
  \label{fig:AUCbleRFkinOF}
\end{minipage}
\end{figure}

\begin{figure}
\centering
\begin{minipage}{.8\textwidth}
  \centering
  \includegraphics[width=\textwidth]{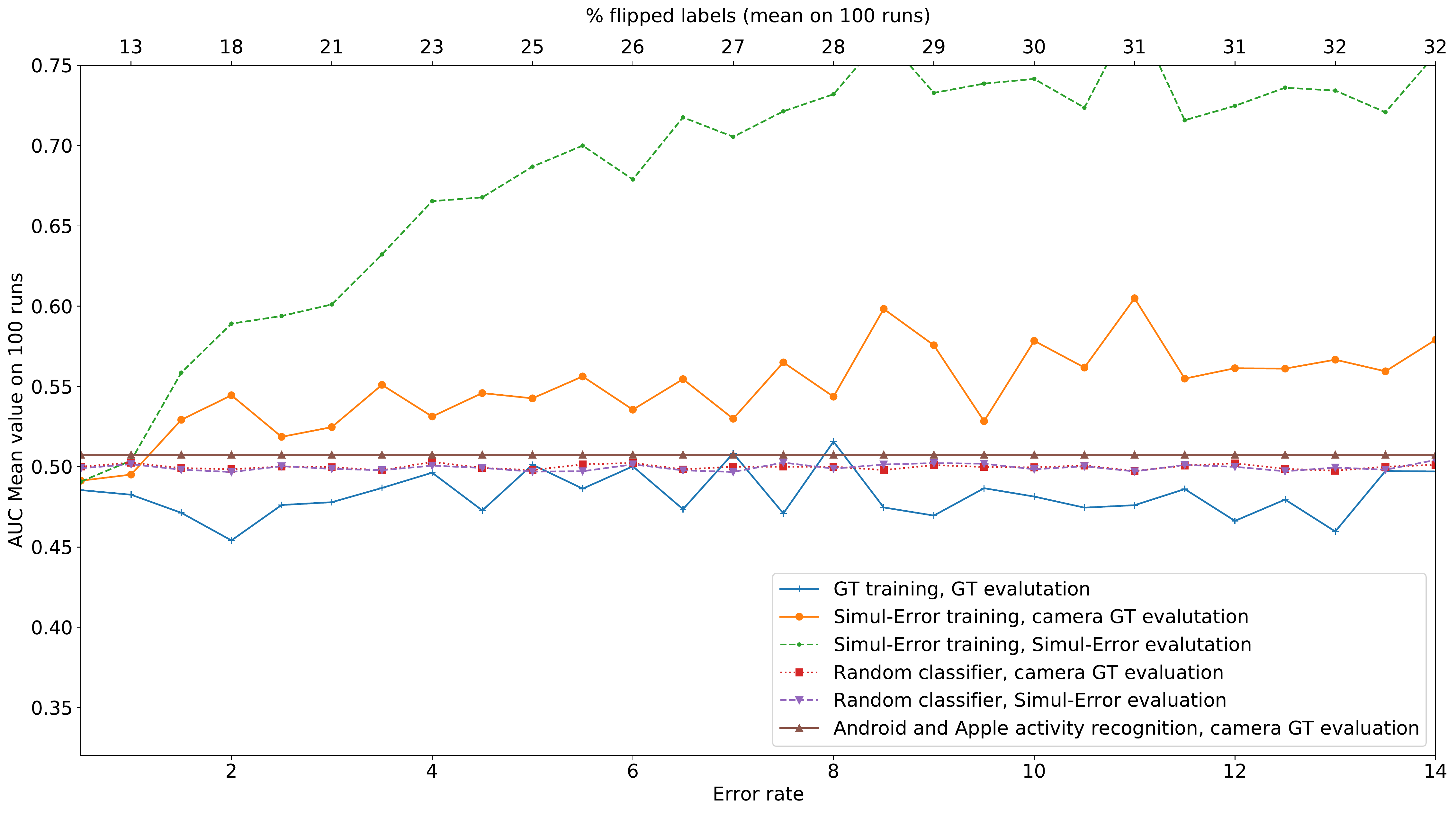}
  \vspace{-3mm}
  \caption{Multi Layer Perceptron one flip experiment
  \newline AUC Classification task using BLE RSSI signal only (p-values $<< 0.01$)
  }
  
  \label{fig:AUCbleNNixxOF}
\end{minipage}%

\begin{minipage}{.8\textwidth}
  \centering
  \includegraphics[width=\textwidth]{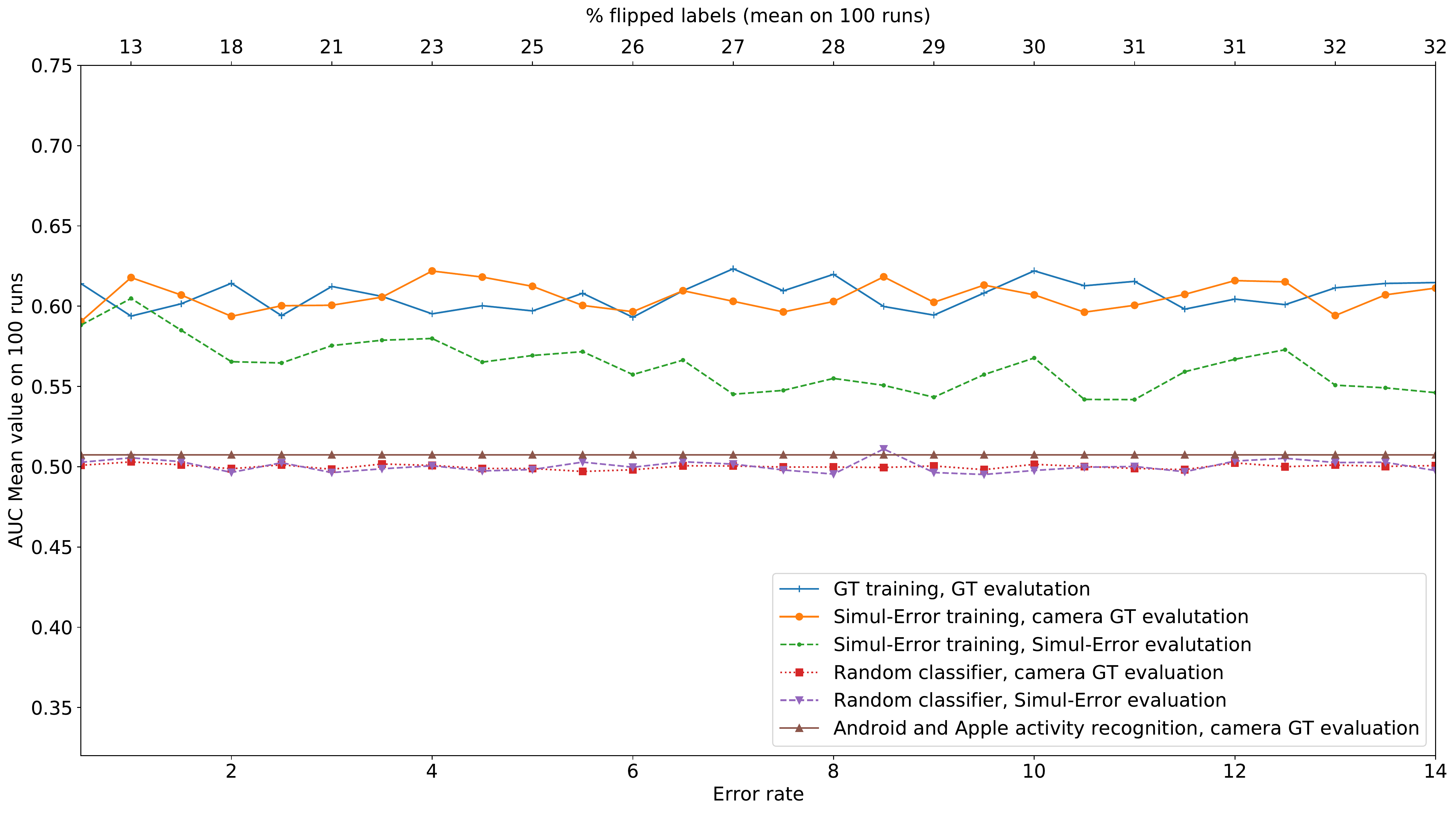}%
  \vspace{-3mm}
  \caption{Multi Layer Perceptron one flip experiment
  \newline
  AUC Classification task using GPS signal only (p-values $<< 0.01$)
  }
  
  \label{fig:AUCbleNNkinOF}
\end{minipage}
\end{figure}

\begin{figure}
\centering
\begin{minipage}{.8\textwidth}
  \centering
  \includegraphics[width=\textwidth]{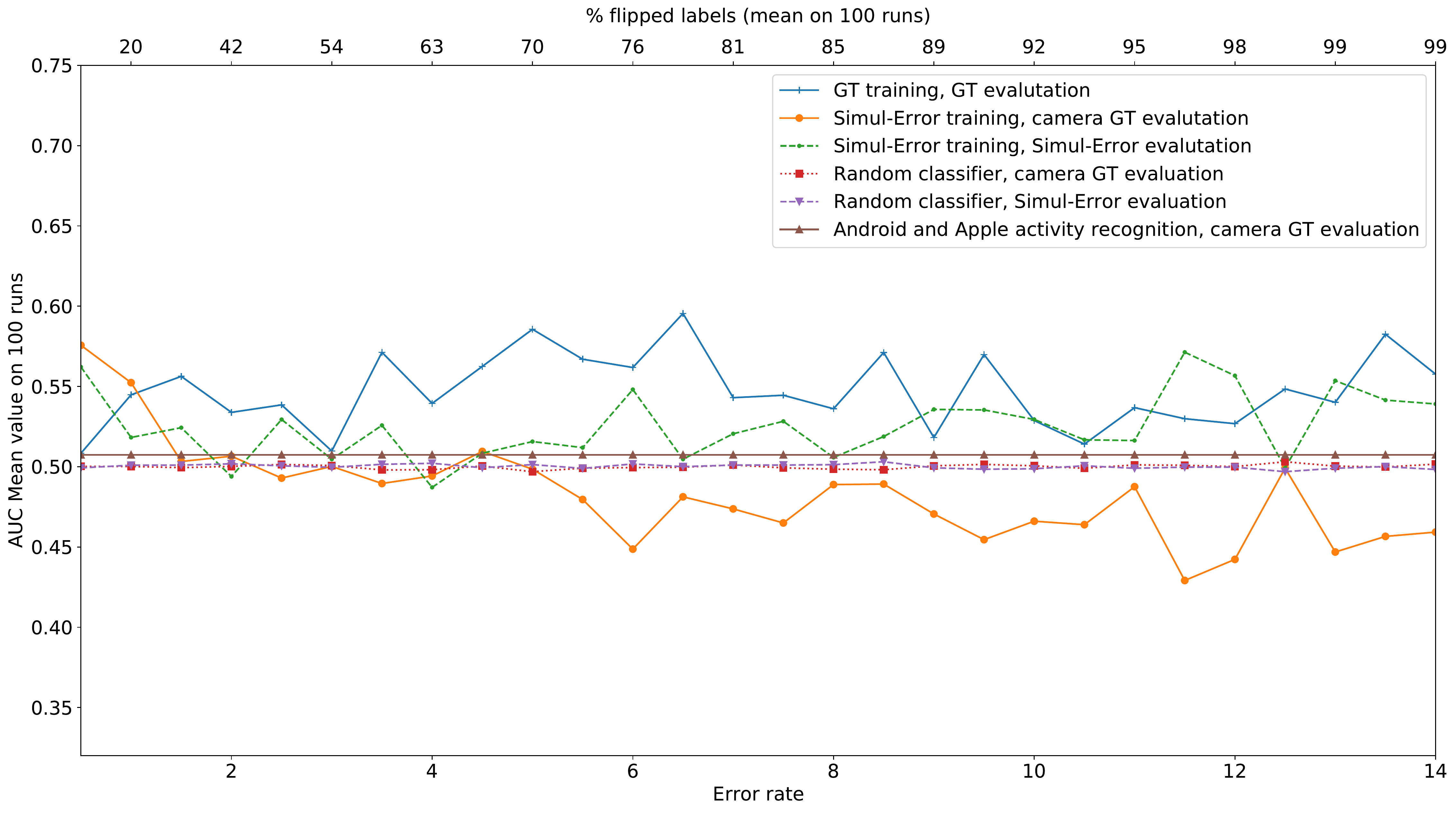}
  \vspace{-3mm}
  \caption{Random Forest two flip experiment
  \newline AUC Classification task using BLE RSSI signal only (p-values $<< 0.01$)
  }
  
  \label{fig:AUCbleRFixxFF}
\end{minipage}%

\begin{minipage}{.8\textwidth}
  \centering
  \includegraphics[width=\textwidth]{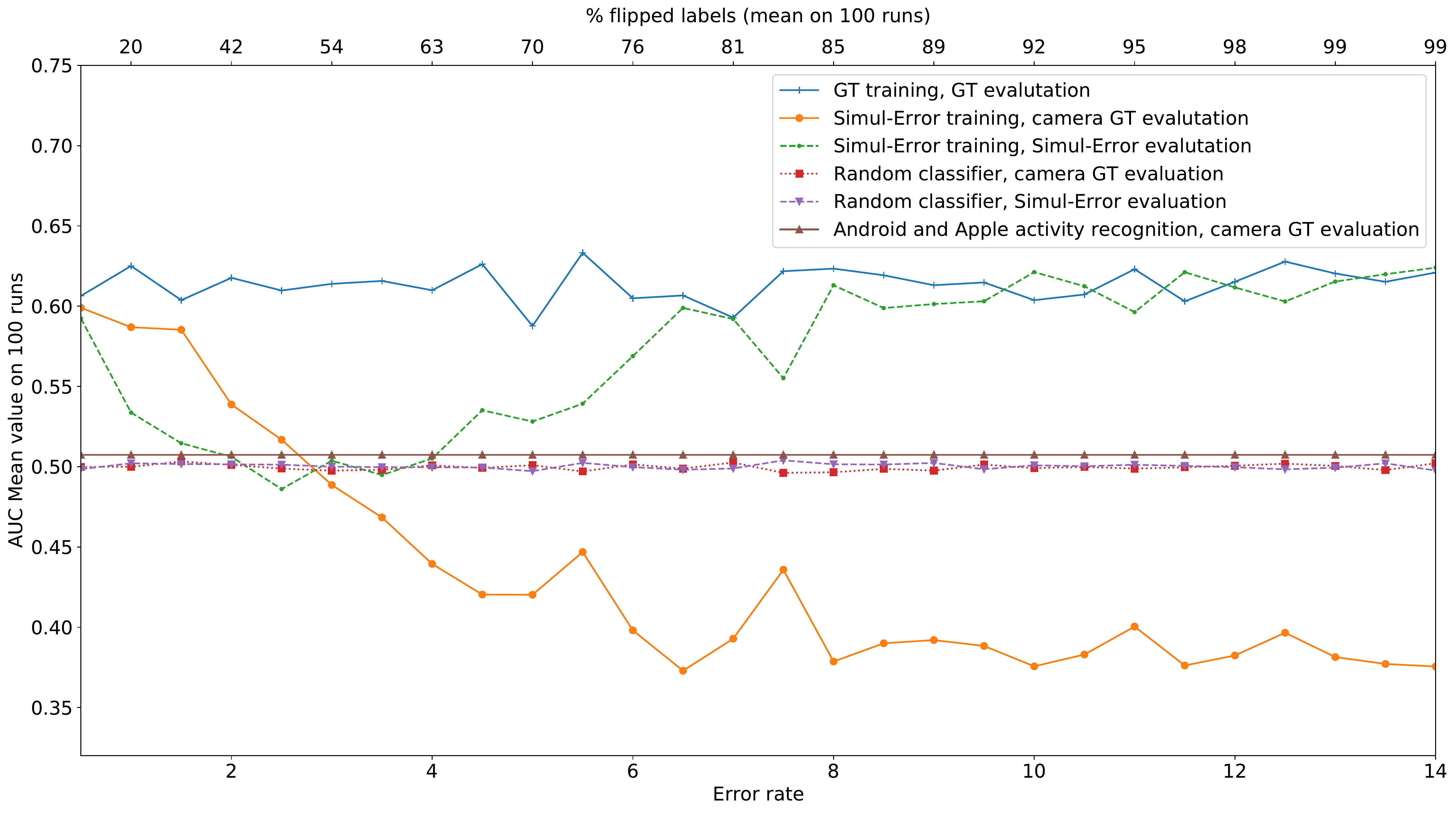}%
  \vspace{-3mm}
  \caption{Random Forest two flip experiment 
  \newline
  AUC Classification task using GPS signal only (p-values $<< 0.01$)
  }
  
  \label{fig:AUCbleRFkinFF}
\end{minipage}
\end{figure}

\begin{figure}
\centering
\begin{minipage}{.8\textwidth}
  \centering
  \includegraphics[width=\textwidth]{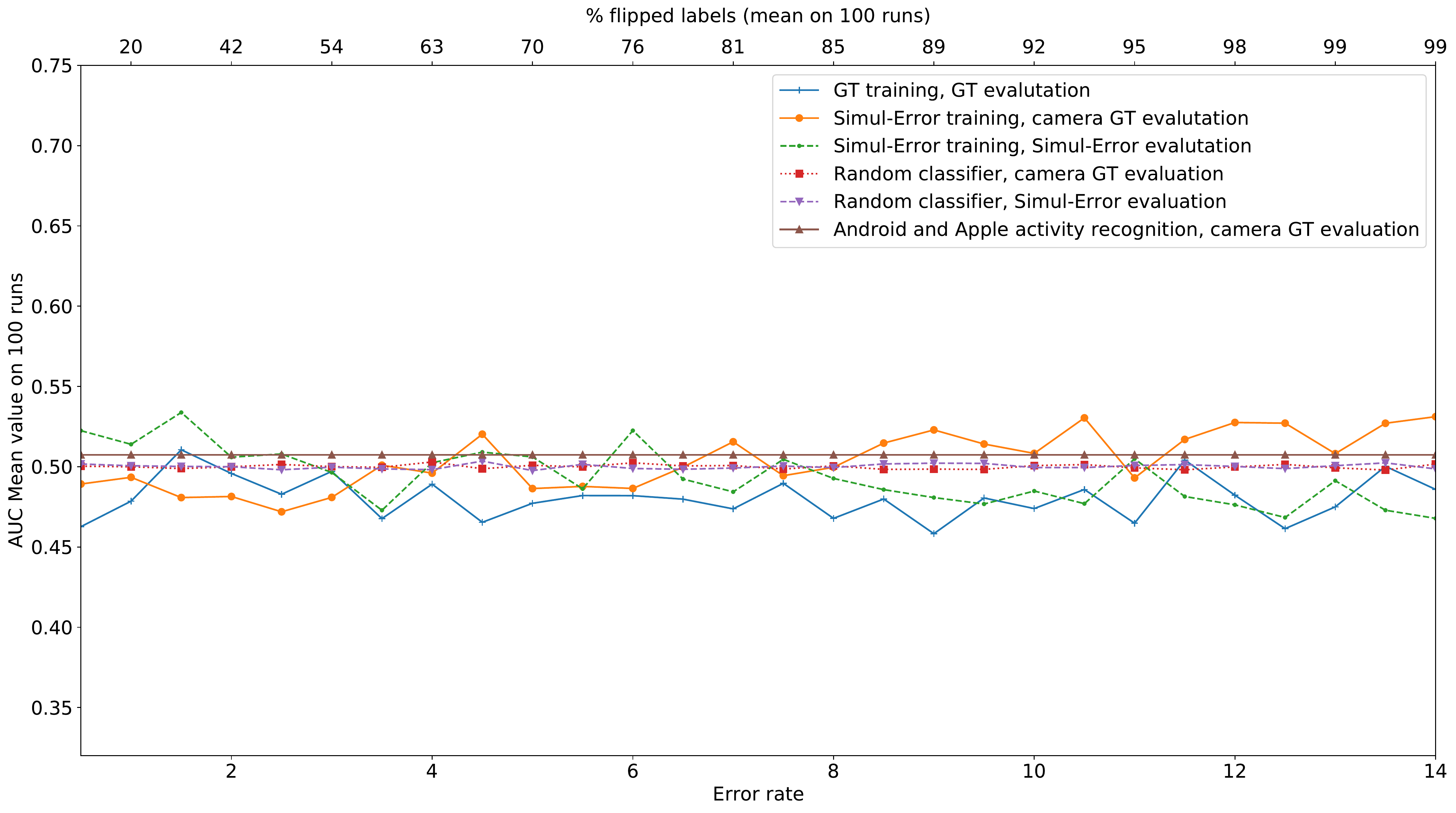}
  \vspace{-3mm}
  \caption{Multi Layer Perceptron two flip experiment
  \newline AUC Classification task using BLE RSSI signal only (p-values $<< 0.01$)
  }
  
  \label{fig:AUCbleNNixxFF}
\end{minipage}%

\begin{minipage}{.8\textwidth}
  \centering
  \includegraphics[width=\textwidth]{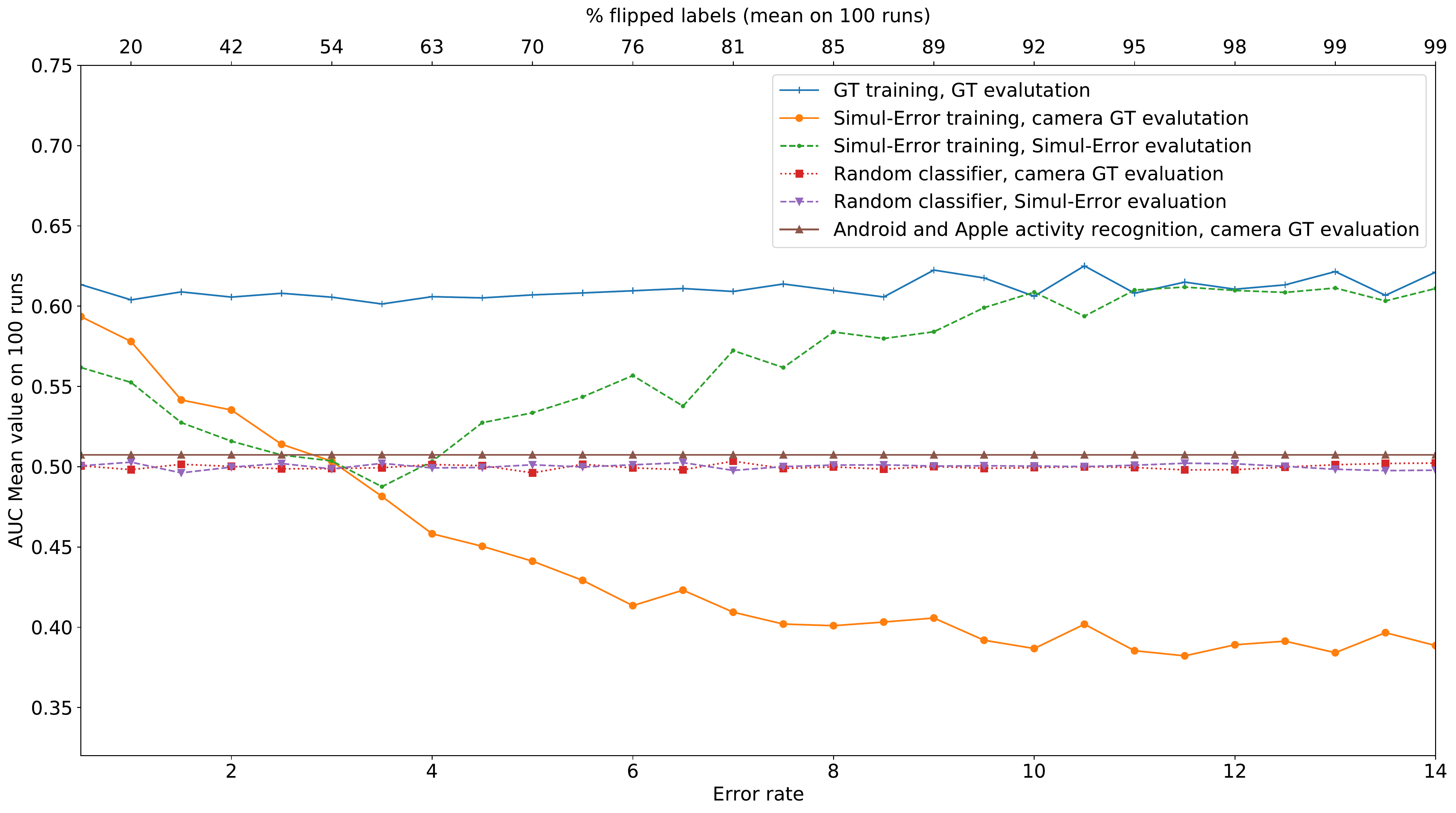}%
  \vspace{-3mm}
  \caption{Multi Layer Perceptron two flip experiment
  \newline
  AUC Classification task using GPS signal only (p-values $<< 0.01$)}
  
  \label{fig:AUCbleNNkinFF}
\end{minipage}
\end{figure}

We repeated classifier training and evaluation with different error levels. Sampling from the same Poisson distribution we propagated errors to the ground-truth under two labeling-flip assumptions. In the first assumption, wrong users' counts will cause some segments to flip their correct class and match the previous or following segment's label. This assumption is consistent with the experiment setup. In the second assumption, more general, the discrete number of errors sampled from Poisson propagates to a random sample of trip segments by flipping the label from the correct class to the alternative. Under the first assumption, Figs. \ref{fig:AUCbleRFixxOF} and  \ref{fig:AUCbleNNixxOF} show the AUC performance of each classifier using BLE beacons against the smartphones OS activity recognition, and the random classifier, for various errors levels. Similarly, Figs. \ref{fig:AUCbleRFkinOF} and  \ref{fig:AUCbleNNkinOF} show the performance of each classifier using GPS sensors. Under the second assumption,  Figs. \ref{fig:AUCbleRFixxFF} and \ref{fig:AUCbleNNixxFF} show the AUC performance using BLE; Figs. \ref{fig:AUCbleRFkinFF} and  \ref{fig:AUCbleNNkinFF}, using GPS.

When evaluating these classifiers on camera-ground-truth, after training on flipped labels at various rates, results suggest that RF are more sensitive to noisy labels when processing GPS features than when processing BLE features. The effect of noisy labels on multi-layer perceptron is negligible when processing GPS features; results show some slight performance improvements when errors propagate according to the first assumption. 

We also note that in any case where high-quality ground-truth is not available, despite the "true" and "unknown" performance of these classifiers, the score is somewhat strongly biased, at a different rate according to the classifier. However, 
BLE signal combined with GPS and other sensors seem to have the potential of improving hybrid BIBO systems, more accurate and less energy-intensive.

Finally, we highlight that dealing with RSSI signal in the experimental context was challenging, and further work could enhance the process of feature extraction from such a weak signal.

\begin{table}[ht]
    \centering
    \caption{Features~\citep{lubba2019catch22} extracted from sensors' signals, within 10 seconds moving window: BLE RSSI and GPS Speed, Space- and Time-gap}
    \begin{tabular*}{\textwidth}{r|l
    }
         \hline 
         1 &  
         Mean value\\
         2& 
         Max value\\
         3& 
         Min value\\
         4& 
         Position where the minimum value is located\\
         5& 
         Position where the maximum value is located\\
         6&
         Amplitude between min and max value\\
         7& Number of points beyond one standard deviation\\
         8& Number of points below one standard deviation\\
         9& Number of points above one standard deviation\\
         10& Number of peaks in 10 seconds window\\
         11& Number of peaks 5 seconds window\\
         12& Number of peaks above 1 standard deviation\\
         13& Peak distance within the same time window\\
         14& Slope\\
         \hline
    \end{tabular*}

    \label{tab:fearCorr}
\end{table}

%

\vspace{-5mm}
\section{Conclusion}
This paper investigates the realistic large-scale deployment of a BIBO system based on BLE beacons, and analyzes the sensitivity of its ML components to errors on labels. The experimental setup recreates challenging conditions with a high density of BLE signals and low speeds of both users and vehicles present in the transport network. 

We test our hypotheses on Poisson error distribution characterizing the labels collection process with person-to-device interactions, typical of current smartphone-based travel surveys. We find that users' validation errors affect both wrong and correct predictions. In the first case users are often unable to correct all the errors. In the second case, users introduce errors by amending correct predictions. Overall, users' did not improve significantly the ground-truth quality. Consequently, data cleansing process should take this factor into consideration beforehand. 

We evaluated RF and MLP, first on BLE beacons signal and second on GPS. In addition, we evaluated the native Android and iOS classifiers, which rely mostly on the accelerometer. These classifiers comparison is based on AUC metric, which assigns the same score, $0.5$, to both random classifiers and classifiers predicting one class only.

We find that off-the-shelf classifiers, based on the accelerometer, perform very close to the random classifier in this experimental context. 
Classifiers based on BLE beacons and GPS perform significantly better when trained with high-quality labels in the same context. When trained on noisy labels and evaluated on high-quality labels, at different levels, MLP seems more robust than RF on GPS features. Yet, when processing BLE features, MLP performance is below the random classifier. 
Overall, Random Forest performs significantly better on both BLE and GPS. At the same time, RF proves to be also robust to noise more on GPS than BLE. 

When high-quality labels are unavailable---even when the noise rate is relatively low---and classifiers are trained and evaluated blindly, the classifiers' evaluation yields a significant and large bias level, underestimating or overestimating the real performance. This problem may affect many results in the available literature. Therefore, efforts will be directed in two directions. One for supervised classifiers such as RF, towards the development of methodologies to assess performance sensitivity on labels' quality. Another for MLP and neural networks, towards architectures able to reduce or eliminate the dependency from labels. 

\section{Imputation}
\label{sec:imputation}
\begin{figure}[!t]
\centering
\includegraphics[width=0.8\textwidth]{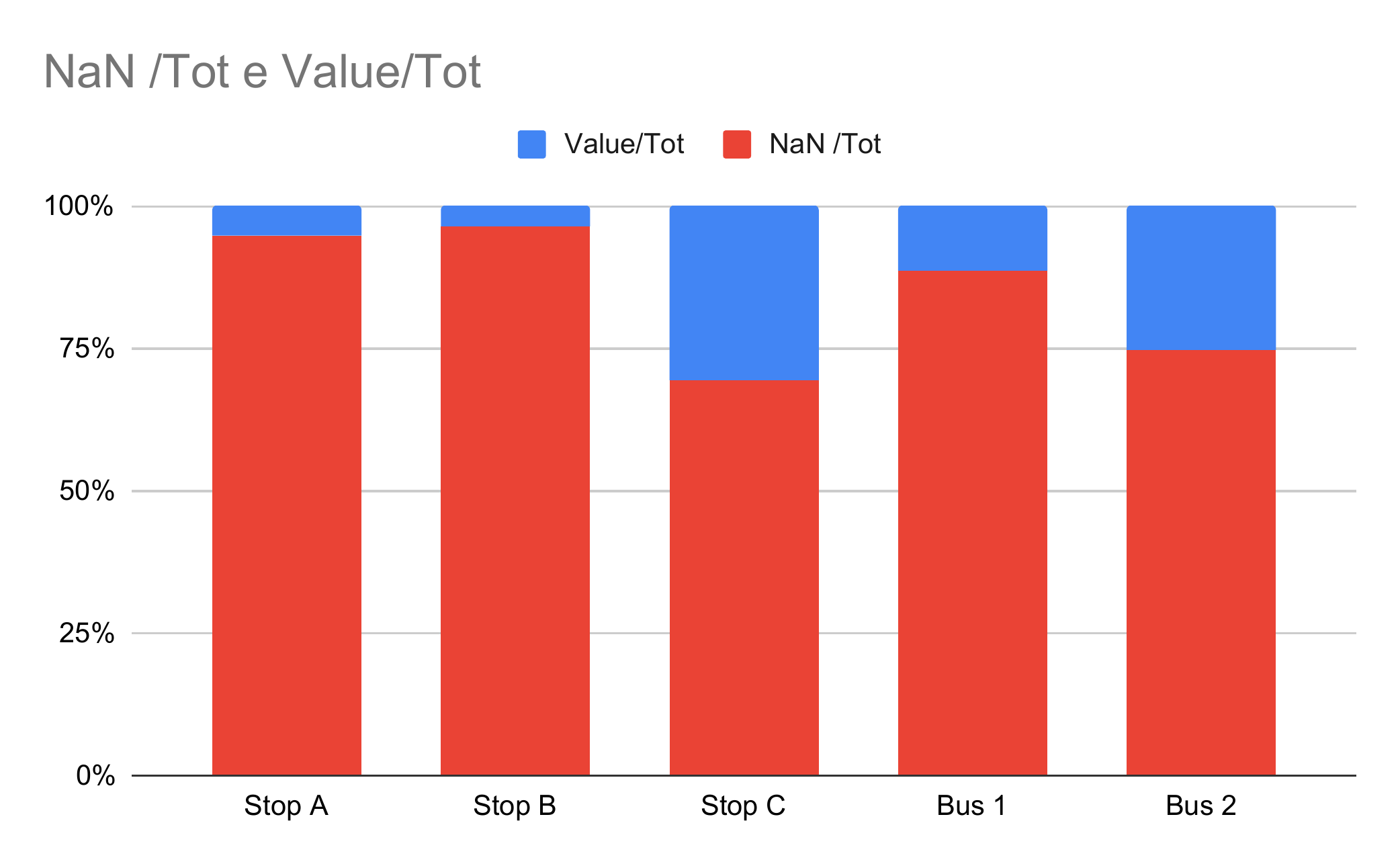}%
\caption{Beacons RSSI timestamp with values vs. Not a Number (NaN), on total points available.}
\label{fig:NaNvsTOT}
\end{figure}

Fig. \ref{fig:NaNvsTOT} shows that compared to the total points collected by the sensing platform, BLE beacons readings are present only on a fraction of the points where GPS is present; the rest of the points are empty. BLE signal goes undetected when the receiver device is not in the beacon range. However, the relative position of the two devices to the user's body often leads to the same result even when the two are in range~\citep{li2017a}. Therefore, we need to perform imputation and fill the gaps whenever appropriate. Existing work shows multiple techniques. Although Kalman-filters might seem the obvious choice from indoor experience~\citep{chen2015a}, this use case requires a faster and relatively more trivial method. From this standpoint, we consider 
exponential-weighted-moving-average (EWMA), 
which consists of computing the average of the readings within a time window, where points close to the center window have a higher weight than points at the end of the window \citep{Shu2014}. 
For EWMA, the weight depends on the window size and the decay rate. From the perspective of a fingerprinting approach (see Sec. \ref{sec:relatedWork}), especially on large-scale deployments, we need to inform the classifier on points where the imputation algorithm could not fill the gaps. We cannot use zero, because BLE beacons signal domain can be found, empirically, in the following domain $RSSI \in (-100,-50)$ \citep{Paek2016}. Further, smartphones record the null value when on the fringe of a BLE range, which is counter-intuitive given the signal's domain. Therefore, because of the meaning of null value and the expected amount of gaps, filling these gaps with zero is likely to poison any classifier. Instead, we can fill these positions with an arbitrary constant and augment the fingerprint vector reporting a weight $0$ in the position filled with the arbitrary constant and $1$ otherwise \citep{Malmberg_2014}. We call this step imputation trick. \eqref{eq:RSSIgaps} defines the fingerprint vector at time $t$ as $FP_t$, where $v_i$ represents the RSSI signal received from the $i^{th}$ BLE beacon, while $v_{j_{GAP}}$ is the gap of signal from the $j^{th}$ BLE beacon. To account for gaps during the learning process, and avoid poisoning the classifier, $FP_t \in {R^{m+n}}$ can be augmented, resulting in a new vector $FPA_t \in {R^{2\cdot(m+n)}}$ \eqref{eq:imputationTrick}, where $v_{j_{IMP}}$ correspond to the signal imputation of the $j^{th}$ BLE beacon gap, for example with EWMA, while $v_{k_{CONST}}$ represent the remaining signal gap from the $j^{th}$ BLE beacon, filled with an arbitrary not null constant. We pass this information to the classifier through the aforementioned binary part of the augmented vector $FPA_t$.

\begin{equation}
\label{eq:RSSIgaps}
\begin{split}
    FP_t = & (v_0, v_1, \dots, v_{0_{GAP}}, \\
    & \dots, v_i, \dots, v_{n_{GAP}}, \\
    & \dots, v_m), \\
    & m > 0 \land n \geq 0 \land i \in (1,m) \\
\end{split}
\end{equation}

\begin{equation}
\label{eq:imputationTrick}
\begin{split}
FPA_t = & (v_0, v_1, \dots,\\ 
           & v_{0_{IMP}}, \dots , v_i, \dots , v_{n_{IMP}}, \dots,\\ 
           & v_{0_{CONST}}, \dots , v_j, \dots , v_{k_{CONST}},\\
           & \dots , v_m,\\
           & 1_0, 1_1, \dots\\ 
           & 1_0, \dots , 1_i, \dots , 1_n, \dots, \\
           & 0_0, \dots , 1_j, \dots , 0_k \\
           & \dots , 1_m), \\
           & m>0 \land n\geq0 \land k\geq0 \\
           & \land i,j \in (1,m), i \ne j \\
           & v_{s_{CONST}} = v_{p_{CONST}} = C, \\
           & \forall s,p \in [0,k], s \ne p
\end{split}
\end{equation}
\section{Algorithms}
This section lists the pseudo-code of the algorithms implemented for the simulation. Alg. \ref{alg:prepData} refers to the data preparation and Alg. \ref{alg:errorprop} to the error simulation and propagation. Alg. \ref{alg:montecarlo} encompasses both Alg. \ref{alg:prepData} and \ref{alg:errorprop}, and performs the following steps.
\begin{enumerate}
    \item Iterative grid-search of the optimal hyperparameteres, accomplished only once per setting, at loop $C=0$.
    \item Model Training, accomplished at each loop $C \geq 0$, using the same optimal hyperparameters found at loop $C=0$.
    \item Model Evaluation, accomplished the four settings of interest.
    
\end{enumerate} 

These four settings of interest are the following. 
\begin{enumerate}
        \item evaluation on camera GT of the model trained with camera GT;
        \item evaluation on GT with flipped labels of the model trained on GT with flipped labels;
        \item evaluation on camera GT of the model trained on GT with flipped labels.
        \item evaluation of a random classifier on camera GT.
\end{enumerate}
\begin{algorithm}[]
\label{alg:prepData}

\DontPrintSemicolon
\SetAlgoLined

\BlankLine

\KwResult{Clean trajectories, assign trip IDs, and extract standardized features for both GPS and BLE signals}

\BlankLine

\SetKwInOut{Input}{Input}\SetKwInOut{Output}{Output}
\Input{raw dataset (RD), true labels (TL)}

\BlankLine

\Output{dataset with tripID labels and features vectors (CD)}
\BlankLine
\BlankLine
$UULIST\leftarrow$ list-unique-users(RD)

\ForEach{user $\in$ UULIST}{
$tripIDs_{\text{user}}\leftarrow$ clean-segment-trajectories(RD, user, TL)
    
    \ForEach{TS $\in$ $\{BLE,GPS\}$}{
    \If{TS == BLE}{
    $CD_{user}\leftarrow$ imputation-trick(D, user, TS, $tripIDs_{\text{user}}$)
    }
    
    $CD_{user}\leftarrow$ extract-standard-features($CD_{user}$, TS)
    }
    CD.insert($CD_{user}$)
}
\Return CD

\caption{Data preparation}
\end{algorithm}
\newpage
\begin{algorithm}
\label{alg:errorprop}
\smaller
\DontPrintSemicolon
\SetAlgoLined

\BlankLine

\KwResult{Faulty ground-truth vector}

\BlankLine

\SetKwInOut{Input}{Input}\SetKwInOut{Output}{Output}
\Input{true labels vector (TL), unique users list (UULIST), features-from-pre-processed-dataset (FCD, see Alg. \ref{alg:prepData})}

\BlankLine

\Output{flipped labels vector FL}
\BlankLine

\ForEach{user $\in$ UULIST}{
            \tcc{draw errors number from Poisson distribution}
            $NE\leftarrow$ draw-from-Poisson(ERR)    \;
            \tcc{Draw NE random TripIDs, as mislabeled trips}
            $\text{WrongTID}_{user}\leftarrow$ draw-random-tripIDs(NE, ${FCD}_{user}$) \;

            \tcc{Copy TL and flip labels for each trip drawn in the previous step}
            
            $FL\leftarrow$ TL \;
            
            \ForEach{trip $\in$ $\text{WrongTID}_{user}$}{
                $FL_{trip}\leftarrow$ flip-labels($FL_{trip}$) \;
            }
        }
\Return FL
\caption{Simulate and propagate P2D validation errors}
\end{algorithm}
\newpage
\begin{algorithm}
\label{alg:montecarlo}
\tiny
\DontPrintSemicolon
\SetAlgoLined

\BlankLine

\KwResult{BIBO Performance distributions of RF and MLP models, evaluated separately for BLE and GPS features, over different average error rates}

\BlankLine

\SetKwInOut{Input}{Input}\SetKwInOut{Output}{Output}
\Input{features-from-pre-processed-dataset (FCD, see Alg. \ref{alg:prepData}), true-labels (TL), target-signal (TS), 
hyperparameters-search-space (HSS, see Table \ref{tab:RFparametersSet}, \ref{tab:MLPparametersSet}), maximum-error-rate (MERR)}

\BlankLine

\Output{
        F1~\eqref{eq:f1-score}, 
        A~\eqref{eq:accuracy}, 
        AUC,
        Optimal Hyperparameters (OP)}
\BlankLine

\tcc{Simulate flipping labels and evaluate model performance against true ground-truth}
$UULIST\leftarrow$ list-unique-users(FCD)

$ERR\leftarrow$ 0.5    \;

\While{ERR $\leq$ MERR}{
    \While{$C<100$}{    
        
        \tcc{Simlulate users errors and propagate through trajectory labels (see Alg. \ref{alg:errorprop})}
        $\text{FL}\leftarrow$ simulate-and-propagate-error(UULIST, TL, FCD)    \;
        
        \tcc{Create Training- and Validation-set, compliant with OOS principle}
        
        $\text{VA}\leftarrow$ pick-random-user-IDs(UULIST, users-num=2)    \;
        $\text{VA}\leftarrow$ extract-features-trajectories-by-user-ID-from-dataset(FCD,$\text{VA}$)    \;
        
        $\text{TR}\leftarrow$ extract-features-trajectories-by-user-ID-from-dataset(FCD,$\text{VA}^\complement$)    \;
        
        
        \tcc{Evaluate classifiers against true and flipped labels (TL Vs. FL)}
        \ForEach{(TR,VA) $\in$ $\{\text{(TR,VA)}_{GPS}, \{\text{(TR,VA)}_{BLE}\}$}{
            \ForEach{model $\in$ $\{RF, MLP\}$}{
                \ForEach{label $\in$ $\{FL, TL\}$}{
                $L\leftarrow$ label \;
                $M\leftarrow$ model \;
                
                \eIf{C=0}{
                \tcc{Hyperparameters 5-fold grid-search on training-set}
                
                $OP_{L}, (
                F1_{{L}_{CV}}, A_{{L}_{CV}}, AUC_{{L}_{CV}})_{M}\leftarrow$ model-5-fold-grid-search(TR,L) \;
                }{
                \tcc{Train a classifier with optimal hyperparameters and labels L}
                $classifier_{L}\leftarrow$ train-model(TR,L,$OP_{L}$) \;
                }

                \tcc{Hold-out evaluation on true labels and validation-set, of a classifier M trained with input-labels L}
                
                $(
                F1_{{L}_{HO}}, A_{{L}_{HO}}, AUC_{{L}_{HO}})_{M}\leftarrow$ evaluate-model($classifier_{L}$, VA, TL)\;
                
                \tcc{Hold-out evaluation on flipped labels and validation-set, of a classifier M trained with input-labels L}
                
                $(
                F1_{{L}_{HO}}, A_{{L}_{HO}}, AUC_{{L}_{HO}})_{M}\leftarrow$ evaluate-model($classifier_{L}$, VA, FL)\;
                
                \tcc{Hold-out evaluation on labels L and validation-set, of random classifier}
                
                $(
                F1_{{RL}_{HO}}, A_{{RL}_{HO}}, AUC_{{RL}_{HO}})_{M}\leftarrow$ evaluate-model($random$, VA, L)\;
                
                $(F1,A,AUC,OP)$.insert$(F1,A,AUC,OP)_{M}$
                }
            }
        }
        $C\leftarrow$ C+1    \;

    }
$ERR\leftarrow$ ERR+0.5    \;
}
\Return(
OP
, F1
, A
, AUC
)
\caption{Model/Sensor performance estimation}
\end{algorithm}
\newpage

\section*{Acknowledgment}
This project is co-financed by the European Regional Development Fund through the Urban Innovative Actions Initiative.





\bibliographystyle{unsrtnat}
\bibliography{bib}
%

%




\end{document}